\documentclass[a4paper]{panl}
\usepackage{cite}
\usepackage{wrapfig}
\usepackage{graphicx}
\usepackage{amssymb}
\usepackage{amsfonts}
\usepackage{amsmath}
\usepackage{longtable}
\usepackage{rotating}
\usepackage{lscape}
\usepackage{epsfig}
\usepackage{multirow}

\originalTeX
\begin{document}

\title{Elastic proton-proton and pion-proton scattering in holographic QCD 
}

\maketitle

\vspace{1mm}

\authors{A.\,Watanabe$^{a,}$\footnote{E-mail: watanabe.akira@oshima-k.ac.jp}}

\from{$^{a}$\,National Institute of Technology, Oshima College, Oshima 742-2193, Japan}

\vspace{3mm}

\begin{abstract}
The elastic proton-proton and pion-proton scattering processes are investigated in the framework of holographic QCD. Considering the Pomeron and Reggeon exchange in the Regge regime, the total and differential cross sections are calculated. In the model setup, the Pomeron and Reggeon exchange are described by the Reggeized spin-2 glueball and vector meson propagator, respectively. For the differential cross sections, contributions of the Coulomb interaction are also taken into account. Adjustable parameters involved in the model are determined with the experimental data, and it is presented that the resulting cross sections are consistent with the data in a wide kinematic region.
\end{abstract}
\vspace*{6pt}


\label{sec:intro}
\section*{Introduction}

Understanding the structure of light hadrons, such as the nucleon and the pion, is one of the most important problems in high energy physics.
To realize it, studies of high energy scattering processes, involving such hadrons, play a crucial role.
In this work, the elastic hadron-hadron scattering processes are investigated in the framework of holographic QCD, focusing on the Regge regime.
Although the elastic hadron-hadron scattering is a simple process, it is almost impossible to compute its cross sections with first-principles calculations due to the nonperturbative nature.
Holographic QCD is an effective approach to the nonperturbative region in QCD.
It enables us to perform the analysis in such a region for various physical quantities.

In this article, we introduce results of our recent analysis on the elastic proton-proton~\cite{Liu:2022zsa,Zhang:2023nsk} and pion-proton~\cite{Liu:2023tjr,Zhang:2024psj} scattering in holographic QCD.
Focusing on the Regge regime, the total and differential cross sections are calculated.
In our model setup, the Pomeron and Reggeon exchange are described by the Reggeized spin-2 glueball and vector meson propagator, respectively.
Furthermore, for the differential cross sections, contributions of the Coulomb interaction are also taken into account.
Adjustable parameters involved in the model are determined with the experimental data of the total cross sections, and then the differential cross sections can be calculated without any additional parameter.
It is presented that the resulting cross sections, both the total and differential ones, are consistent with the data, which shows the predictive power of the present model.

Due to the universality of the Pomeron and Reggeon exchange, this approach may be applied to other high energy scattering processes.
The present framework will further be tested at various experimental facilities in the future.

\label{sec:model_setup}
\section*{Model Setup}

\begin{figure}[t]
\begin{center}
\begin{tabular}{ccccc}
\includegraphics[width=0.35\textwidth]{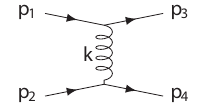}
\includegraphics[width=0.35\textwidth]{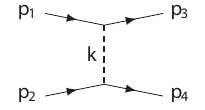}
\end{tabular}
\end{center}
\caption{
The left and right Feynman diagrams represent %
the $pp$($p\bar{p}$) scattering with the Pomeron and Reggeon exchange %
in the $t$-channel, respectively.
}
\label{diagram}
\end{figure}

Based on Ref.~\cite{Liu:2022zsa}, first we introduce the model setup for the proton-proton ($pp$) and proton-antiproton ($p \bar p$) scattering.
We consider the diagrams shown in Fig.~\ref{diagram}.
Taking into account contributions of the Pomeron and Reggeon exchange, which are described by the Reggeized spin-2 glueball and vector meson propagator, the scattering amplitudes are written as
\begin{equation}\label{total_amp_pp}
\mathcal{A}_{\rm tot}^{pp, p \bar{p}} = \mathcal{A}_g^{pp, p \bar{p}} + \mathcal{A}_{v}^{pp, p \bar{p}}.
\end{equation}
The matrix element of the energy momentum tensor $T_{\mu \nu}$ between the initial and final proton states is expressed as
\begin{align}\label{EMT_matrix_element}
\langle p', s' | T_{\mu\nu} | p, s \rangle  = \bar{u} (p', s') \bigg[ &A(t)\frac{\gamma_{\mu} P_{\nu} + \gamma_{\nu} P_{\mu}}{2} \nonumber
\\&+ B(t) \frac{ i (P_{\mu} \sigma_{\nu\rho} + P_{\nu} \sigma_{\mu\rho} ) k^\rho}{4m_p} \nonumber
\\&+ C(t) \frac{k_{\mu} k_{\nu} - \eta_{\mu\nu} k^2}{m_p} \bigg] u (p, s),
\end{align}
where $P = (p_1 + p_3) / 2 $, $k = p_3 - p_1$, $t = - k^2$, and $m_p$ is the proton mass.
$A (t)$, $B (t)$, and $C (t)$ are the proton form factors.
Since we focus on the Regge regime in this study, $A(t)$ gives the dominant contribution and the other two can be neglected.
For $A (t)$, we utilize the proton gravitational form factor, which was derived in Ref.~\cite{Abidin:2009hr} with the bottom-up AdS/QCD model.

The vertex of glueball-proton-proton in the Regge limit is given by
\begin{equation}\label{gpp_vertex}
\Gamma_{g}^{\mu \nu} = \frac{i \lambda_{g} A(t)}{2}(\gamma^\mu P^\nu + \gamma^\nu P^\mu),
\end{equation}
where $\lambda_g$ is the coupling constant. %
The massive spin-2 glueball propagator is written as
\begin{equation}\label{glueball_propagator}
D_{\alpha \beta \gamma \delta}^g (k) = \frac{- i d_{\alpha \beta \gamma \delta}}{k^2 + m_g^2},
\end{equation}
where $m_g$ is the glueball mass, and $d_{\alpha \beta \gamma \delta}$ is expressed as
\begin{align}
d_{\alpha \beta \gamma \delta} = &\frac{1}{2}(\eta_{\alpha \gamma}\eta_{\beta \delta} + \eta_{\alpha \delta} \eta_{\beta \gamma}) - \frac{1}{2m_g^2}(k_\alpha k_\delta \eta_{\beta \gamma} + k_\alpha k_\gamma \eta_{\beta \delta} + k_\beta k_\delta \eta_{\alpha \gamma} + k_\beta k_\gamma \eta_{\alpha \delta}) \nonumber \\
&+ \frac{1}{24} \left[ \left(\frac{k^2}{m_g^2} \right)^2 - 3 \left( \frac{k^2}{m_g^2} \right) - 6 \right] \eta_{\alpha \beta} \eta_{\gamma \delta} - \frac{k^2 - 3m_g^2}{6 m_g^4}(k_\alpha k_\beta \eta_{\gamma \delta} + k_\gamma k_\delta \eta_{\alpha \beta}) \nonumber \\
&+ \frac{2 k_\alpha k_\beta k_\gamma k_\delta}{3 m_g^4}.
\end{align}
In the Regge limit, only the first term in the right-hand side needs to be considered.
Combining Eqs.~\eqref{gpp_vertex} and \eqref{glueball_propagator}, the glueball exchange amplitude is obtained as
\begin{align}\label{glueball_amplitude}
\mathcal{A}_g^{pp (p \bar{p})} &= (\bar{u}_1 \Gamma_g^{\alpha \beta}u_3) D_{\alpha \beta \gamma \delta}^g (k) (\bar{u}_2 \Gamma_g^{\gamma \delta} u_4) \nonumber \\
&= \frac{i \lambda_g^2}{8 (k^2 + m_g^2)}[2 s A^2 (t) (\bar{u}_1 \gamma^\alpha u_3)(\bar{u}_2 \gamma_\alpha u_4) + 4 A^2 (t) p_2^\alpha p_1^\beta (\bar{u}_1 \gamma_\alpha u_3)(\bar{u}_2 \gamma_\beta u_4)].
\end{align}

The vector meson propagator is written as
\begin{equation}
D_{\mu \nu}^v (k) = \frac{i}{k^2 + m_v} \eta_{\mu \nu},
\end{equation}
where $m_{v}$ is the vector meson mass.
The vector-proton-proton vertex is expressed as
\begin{equation}
\Gamma_v^\mu = - i \lambda_v \gamma^\mu,
\end{equation}
where $\lambda_v$ is the coupling constant.
These lead to the vector meson exchange amplitude,
\begin{align}\label{vector_meson_amplitude}
\mathcal{A}_v^{pp (p \bar{p})} &= (\bar{u}_1 \Gamma_v^\mu u_3) D_{\mu \nu}^v (k)(\bar{u}_2 \Gamma_v^\nu u_4) \nonumber \\
&= - \frac{i \lambda_v^2}{k^2 + m_v^2} \eta_{\mu \nu}(\bar{u}_1 \gamma^\mu u_3)(\bar{u}_2 \gamma^\nu u_4).
\end{align}

From Eq.~\eqref{total_amp_pp}, the total amplitude is obtained as
\begin{align}
\mathcal{A}^{pp (p \bar{p})}_{\mathrm{tot}} = &\frac{- i \lambda_g^2}{8(t - m_g^2)}[2 s A^2 (t)(\bar{u}_1 \gamma^\alpha u_3)(\bar{u}_2 \gamma_\alpha u_4) + 4 A^2 (t) p_2^\alpha p_1^\beta (\bar{u}_1 \gamma_\alpha u_3)(\bar{u}_2 \gamma_\beta u_4)]  \nonumber \\
&+ \frac{i \lambda_v^2}{t - m_v^2} \eta_{\mu \nu}(\bar{u}_1 \gamma^\mu u_3)(\bar{u}_2 \gamma^\nu u_4).
\end{align}
Considering the conditions, $s \gg | t |$, $u_1 \approx u_3$, and $u_2 \approx u_4$, the differential cross section can be obtained as
\begin{align}\label{dcs_pp_1}
\frac{d \sigma}{dt}
= &\frac{1}{16 \pi s^2} | \mathcal{A}^{pp (p \bar{p})}_{\mathrm{tot}} |^2 \nonumber \\
= &\frac{\lambda_g^4 s^2 A^2 (t)}{16 \pi | t - m_g^2 |^2} - \frac{\lambda_g^2 \lambda_v^2 A^2 (t) s}{8 \pi} \bigg[ \frac{1}{(t - m_g^2)^*} \times \frac{1}{t - m_v^2} + \frac{1}{(t - m_v^2)^*} \times \frac{1}{t - m_g^2} \bigg] \nonumber \\
&+ \frac{\lambda_v^4}{4 \pi | t - m_v^2 |^2}.
\end{align}

Since only the lightest states are considered in Eq.~\eqref{dcs_pp_1}, the propagators need to be Reggeized to include the higher spin states.
Following the Reggeization procedure~\cite{Anderson:2016zon}, they are replaced as follows:
\begin{align}
&\frac{1}{t-m_{g}^2} \rightarrow  \frac{\alpha'_{g}}{2}e^{-\frac{i\pi\alpha_{g}(t)}{2}}\left(\frac{\alpha'_{g}s}{2}\right)^{\alpha_{g}(t)-2}\frac{\Gamma\left[3-\frac{\chi_{g}}{2}\right]\Gamma\left[1-\frac{\alpha_{g}(t)}{2}\right]}{\Gamma\left[2-\frac{\chi_{g}}{2}+\frac{\alpha_{g}(t)}{2}\right]}, \\
&\frac{1}{t-m_{v}^2}  \rightarrow  \alpha'_{v}e^{-\frac{i\pi \alpha_{v}(t)}{2}}\sin\left[\frac{\pi\alpha_{v}(t)}{2}\right]\left(\alpha'_{v}s\right)^{\alpha_{v}(t)-1}\Gamma[-\alpha_{v}(t)],
\end{align}
where $\chi_{g} \equiv \alpha_{g} (s) + \alpha_g (u) + \alpha_g (t)$, in which $\alpha_g (x) = \alpha_g (0) + \alpha'_g x$, and $\alpha_v (x) = \alpha_v (0) + \alpha'_v x$.
With the Reggeized propagators, the differential cross section can be expressed as
\begin{align}
 \frac{d \sigma^{pp (p \bar{p})}}{dt} = &\frac{\lambda_g^4 s^2 A^4 (t)}{16 \pi} \left[ \frac{\alpha'_g}{2} \frac{\Gamma \left[3 - \frac{\chi_g}{2} \right] \Gamma \left[1 - \frac{\alpha_g(t)}{2} \right]}{\Gamma \left[2 - \frac{\chi_g}{2} + \frac{\alpha_g (t)}{2} \right]}\left(\frac{\alpha'_g s}{2}\right)^{\alpha_g (t) - 2} \right]^2 \nonumber \\
&- \frac{\lambda_g^2 \lambda_v^2 s A^2 (t)}{4 \pi} \left[ \frac{\alpha'_g}{2} \frac{\Gamma \left[3 - \frac{\chi_g}{2} \right] \Gamma \left[1 - \frac{\alpha_g(t)}{2} \right]}{\Gamma \left[2 - \frac{\chi_g}{2} + \frac{\alpha_g (t)}{2} \right]}\left(\frac{\alpha'_g s}{2} \right)^{\alpha_g (t) - 2}\right] \nonumber \\
&\times\left[\alpha'_v \sin\bigg( \frac{\pi \alpha_v (t)}{2} \bigg) (\alpha'_v s)^{\alpha_v (t) - 1}\Gamma[- \alpha_v (t)] \right] \cos \left[ \frac{\pi}{2}(\alpha_g (t) - \alpha_v (t)) \right] \nonumber \\
&+ \frac{\lambda_v^4}{4 \pi} \left[ \alpha'_v \sin \bigg( \frac{\pi \alpha_v (t)}{2} \bigg) (\alpha'_v s)^{\alpha_v (t) - 1} \Gamma[- \alpha_v (t)] \right]^2,
\label{dcs_pp_final}
\end{align}
which leads to the invariant amplitude,
\begin{align}
\mathcal{A}^{pp (p \bar{p})} (s, t) = &- s \lambda_g^2 A^2 (t) e^{- \frac{i \pi \alpha_g (t)}{2}}\frac{\Gamma \left[3 - \frac{\chi_g}{2} \right] \Gamma \left[1 - \frac{\alpha_g (t)}{2} \right]}{\Gamma \left[2 - \frac{\chi_g}{2} + \frac{\alpha_g (t)}{2}\right]}\left(\frac{\alpha'_g s}{2} \right)^{\alpha_g (t) - 1} \nonumber \\
&+ 2 s \lambda_v^2 \alpha'_v e^{- \frac{i \pi \alpha_v (t)}{2}}\sin \left(\frac{\pi \alpha_v (t)}{2} \right) (\alpha'_v s)^{\alpha_v (t) - 1} \Gamma[- \alpha_v (t)].
\end{align}
Applying the optical theorem, the total cross section is then obtained as
\begin{align}
\sigma^{pp (p \bar{p})}_\mathrm{tot} = &\frac{1}{s} \mathrm{Im} \mathcal{A}^{pp (p \bar{p})} (s, t = 0) \nonumber\\
=&\lambda_g^2 \sin \left(\frac{\pi \alpha_g (0)}{2} \right) \frac{\Gamma\left[3 - \frac{\chi_g}{2} \right] \Gamma \left[1 - \frac{\alpha_g (t)}{2} \right]}{\Gamma \left[2 - \frac{\chi_g}{2} + \frac{\alpha_g (t)}{2} \right]}\left( \frac{\alpha'_g s}{2} \right)^{\alpha_g (0) - 1} \nonumber \\
&- 2 \lambda_v^2 \alpha'_v \sin^2 \left( \frac{\pi \alpha_v (0)}{2} \right) (\alpha'_v s)^{\alpha_v (0) - 1} \Gamma[- \alpha_v (0)].
\label{tcs_pp_final}
\end{align}

For the case of pion-proton ($\pi p$) scattering~\cite{Liu:2023tjr}, the required procedure to obtain analytical expressions for the cross sections is similar to that for the $pp$ case introduced above.
Here, we only mention the difference in the Pomeron-hadron coupling.
The total amplitude for the $\pi p$ scattering is written as
\begin{equation}
\mathcal{A}_{\rm tot}^{\pi p} = \mathcal{A}_{ g}^{\pi p} + \mathcal{A}_{ v}^{\pi p}.
\end{equation}
The glueball-pion-pion vertex can be extracted from the energy momentum tensor matrix element for the pion, which is expanded with two gravitational form factors $A_{\pi} (t)$ and $C_{\pi} (t)$:
\begin{equation}
\langle \pi^a (p_2) | T^{\mu \nu} | \pi^b (p_1) \rangle = \delta^{ab} \bigg[2 A_\pi (t) p^\mu_\pi p^\nu_\pi + \frac{1}{2} C_\pi (t) \big( k_\pi^2 \eta^{\mu \nu} - k^\mu_\pi k^\nu_\pi \big) \bigg],
\end{equation}
where $p_{\pi}$ is the pion four-momentum.
In the Regge limit, the contribution of the $C_\pi (t)$ involved term can be neglected.
To specify $A_{\pi} (t)$, we employ the result obtained in Ref.~\cite{Abidin:2008hn} with the bottom-up AdS/QCD model of mesons.

After Reggeization of the propagators, the invariant amplitude for the elastic $\pi p$ scattering is obtained as
\begin{align}
\mathcal{A}_{\mathrm{tot}}^{\pi p} =
&- \lambda_{g \pi \pi} \lambda_{g p p} A_\pi (t) A_p (t) s^2 \frac{\alpha'_P}{2} e^{- \frac{i \pi \alpha_P (t)}{2}}\bigg( \frac{\alpha'_P s}{2} \bigg)^{\alpha_P (t) - 2}\frac{\Gamma \big[ 3 - \frac{\chi}{2} \big] \Gamma \big[1 - \frac{\alpha_P (t)}{2} \big]}{\Gamma \big[ 2 - \frac{\chi}{2} + \frac{\alpha_P (t)}{2} \big]} \nonumber \\
&+ 2 \lambda_{v \pi \pi} \lambda_{v p p} s \alpha'_R e^{- \frac{i \pi \alpha_R (t)}{2}} \sin \bigg( \frac{\pi \alpha_R (t)}{2} \bigg) (\alpha'_R s)^{\alpha_R (t) - 1}\Gamma[-\alpha_R (t)].
\end{align}
Then, the differential cross section is obtained as
\begin{align}
\frac{d \sigma^{\pi p}}{dt} = &\frac{1}{16 \pi s^2}|\mathcal{A}_{\mathrm{tot}}^{\pi p}|^2 \nonumber \\
= &\frac{1}{16 \pi}\big( \lambda_{g \pi \pi}\lambda_{g p p} A_\pi (t) A_p (t) s \big)^2~\bigg[\frac{\alpha'_P}{2}\bigg(\frac{\alpha'_P s}{2} \bigg)^{\alpha_P (t) - 2}\frac{\Gamma \big[ 3 - \frac{\chi}{2} \big] \Gamma \big[1 - \frac{\alpha_P (t)}{2} \big]}{\Gamma \big[2 - \frac{\chi}{2} + \frac{\alpha_P (t)}{2}\big]}\bigg]^2 \nonumber \\
&+ \frac{1}{4 \pi}(\lambda_{v \pi \pi} \lambda_{v p p})^2~\bigg[\alpha'_R \sin \bigg(\frac{\pi \alpha_R (t)}{2}\bigg) (\alpha'_R s)^{\alpha_R (t) - 1}\Gamma[-\alpha_R (t)]\bigg]^2 \nonumber \\
&- \frac{1}{8 \pi}\big(\lambda_{g \pi \pi}\lambda_{g p p} \lambda_{v \pi \pi}\lambda_{v p p} A_\pi (t) A_p (t) s \big) \nonumber \\
&\times \bigg[\frac{\alpha'_P}{2} e^{-\frac{i \pi \alpha_P (t)}{2}}\bigg(\frac{\alpha'_P s}{2}\bigg)^{\alpha_P (t) - 2}\frac{\Gamma\big[3 - \frac{\chi}{2}\big]\Gamma \big[1 - \frac{\alpha_P (t)}{2} \big]}{\Gamma \big[2 - \frac{\chi}{2} + \frac{\alpha_P (t)}{2}\big]}~, \nonumber \\
&~~~~~~ \alpha'_v e^{-\frac{i \pi \alpha_R (t)}{2}} \sin \bigg(\frac{\pi \alpha_R (t)}{2}\bigg) (\alpha'_R s)^{\alpha_R (t) - 1} \Gamma[- \alpha_R (t)]\bigg]^*,
\end{align}
where $[x,y]^*=xy^*+x^*y$.
Applying the optical theorem, the total cross section can be obtained as
\begin{align}
\sigma^{\pi p}_{\mathrm{tot}} =
&\frac{1}{s}\mathrm{Im}{\mathcal{A}^{\pi p}_{\mathrm{tot}}(s, t = 0)} \nonumber \\
= &\lambda_{g \pi \pi}\lambda_{g p p}\frac{\Gamma\big[3 - \frac{\chi}{2}\big]\Gamma\big[1 - \frac{\alpha_P (t)}{2}\big]}{\Gamma\big[2 - \frac{\chi}{2} + \frac{\alpha_P (t)}{2}\big]}\bigg(\frac{\alpha'_P s}{2}\bigg)^{\alpha_P (0) - 1} \sin \bigg(\frac{\pi\alpha_P (0)}{2}\bigg) \nonumber \\
&- 2 \lambda_{v \pi \pi}\lambda_{v p p}\alpha'_R \sin^2 \bigg(\frac{\pi\alpha_R (0)}{2}\bigg)(\alpha'_R s)^{\alpha_R (0) - 1}\Gamma[- \alpha_R (0)].
\end{align}

Finally, we briefly introduce how to add the Coulomb interaction contribution to the expressions discussed in the earlier part.
The details are presented in Ref.~\cite{Zhang:2023nsk} for the $pp$ and $p \bar p$ scattering and in Ref.~\cite{Zhang:2024psj} for the $\pi p$ scattering.
The total scattering amplitude, including the Coulomb interaction contribution, can be expressed as~\cite{Bethe:1958zz} 
\begin{equation}
F_{N+C} (s, t) = F_N (s, t) + e^{i \alpha \phi (s, t)} F_C (s, t), \label{total_amp_NC}
\end{equation}
where $F_N$ is the strong interaction amplitude, $F_C$ is the Coulomb interaction amplitude, and $\phi$ is the Coulomb phase.
The Coulomb amplitude for point-like charges is expressed with the effective electromagnetic form factor as
\begin{equation}
F_C (s, t) = \mp\frac{8 \pi \alpha s}{| t |} G_{\mathrm{eff}}^2 (t),
\end{equation}
where $\alpha$ is the fine structure constant.
The negative (positive) sign corresponds to the scattering of particles with identical (opposite) charges.
For the Coulomb phase, we employ the result obtained in the eikonal model~\cite{Cahn:1982nr}:
\begin{align}
&\phi = \mp \left[ \ln \left(\frac{B | t |}{2} \right) + \gamma + C \right], \\
&C =\ln \left(1 + \frac{8}{B \Lambda^2} \right) + \left(4 | t | / \Lambda^2 \right) \ln \left(4 | t | / \Lambda^2 \right) + 2 | t | / \Lambda^2,
\end{align}
where $B$ is the diffractive slope at $t \to 0$, $\gamma$ is the Euler constant, and $\Lambda^2$ is a parameter.

\label{sec:numerical_results}
\section*{Numerical Results}

\begin{figure}[!tb]
\centering
\includegraphics[width=0.65\textwidth]{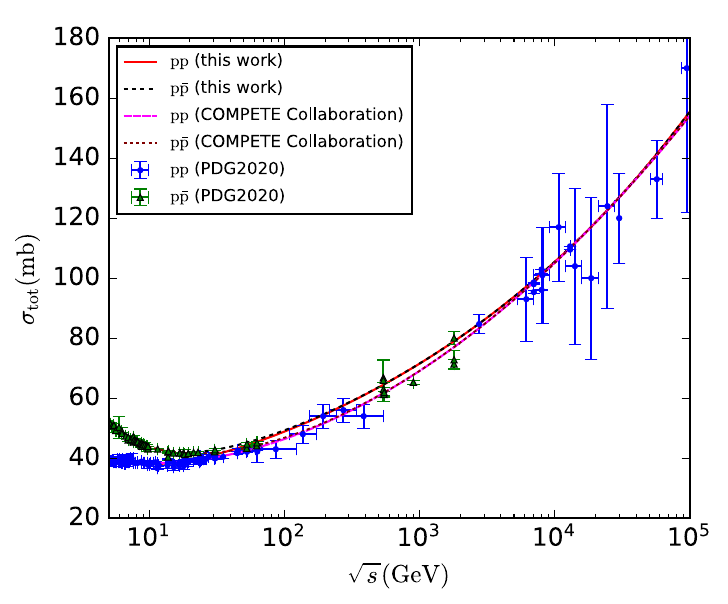}
\caption{
The total cross section of the $pp$ and $p\bar{p}$ scattering %
as a function of $\sqrt{s}$. %
Our calculations are compared with the results obtained %
by the COMPETE Collaboration~\cite{COMPETE:2002jcr}.
The experimental data are %
taken from Ref.~\cite{ParticleDataGroup:2020ssz}.
}
\label{TCS_pp}
\end{figure}

\begin{figure}[!tb]
\centering
\includegraphics[width=0.65\textwidth]{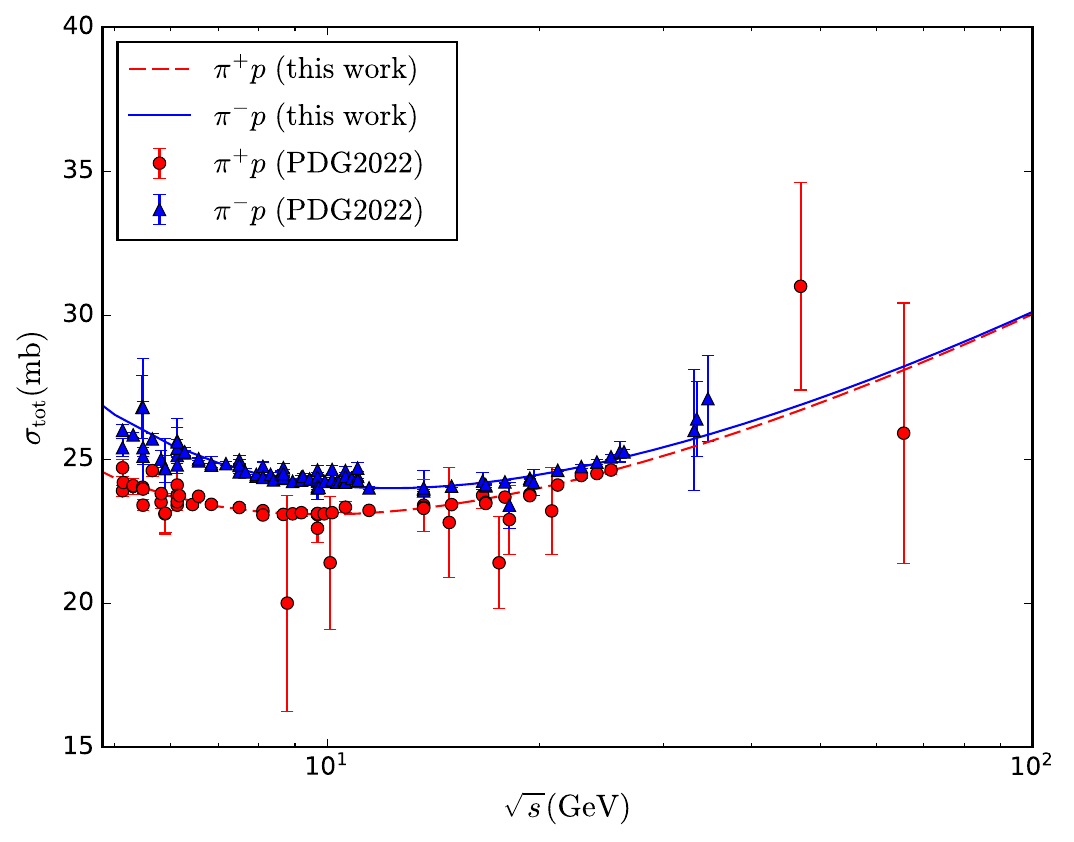}
\caption{
The total cross section of the $\pi p$ scattering as a function of $\sqrt{s}$.
The solid and dashed curves represent our calculations %
for the $\pi^- p$ and $\pi^+ p$ scattering, respectively.
The experimental data are %
taken from Ref.~\cite{ParticleDataGroup:2022pth}.
}
\label{TCS_pip}
\end{figure}

\begin{figure}[!tb]
\centering
\includegraphics[width=1.0\textwidth]{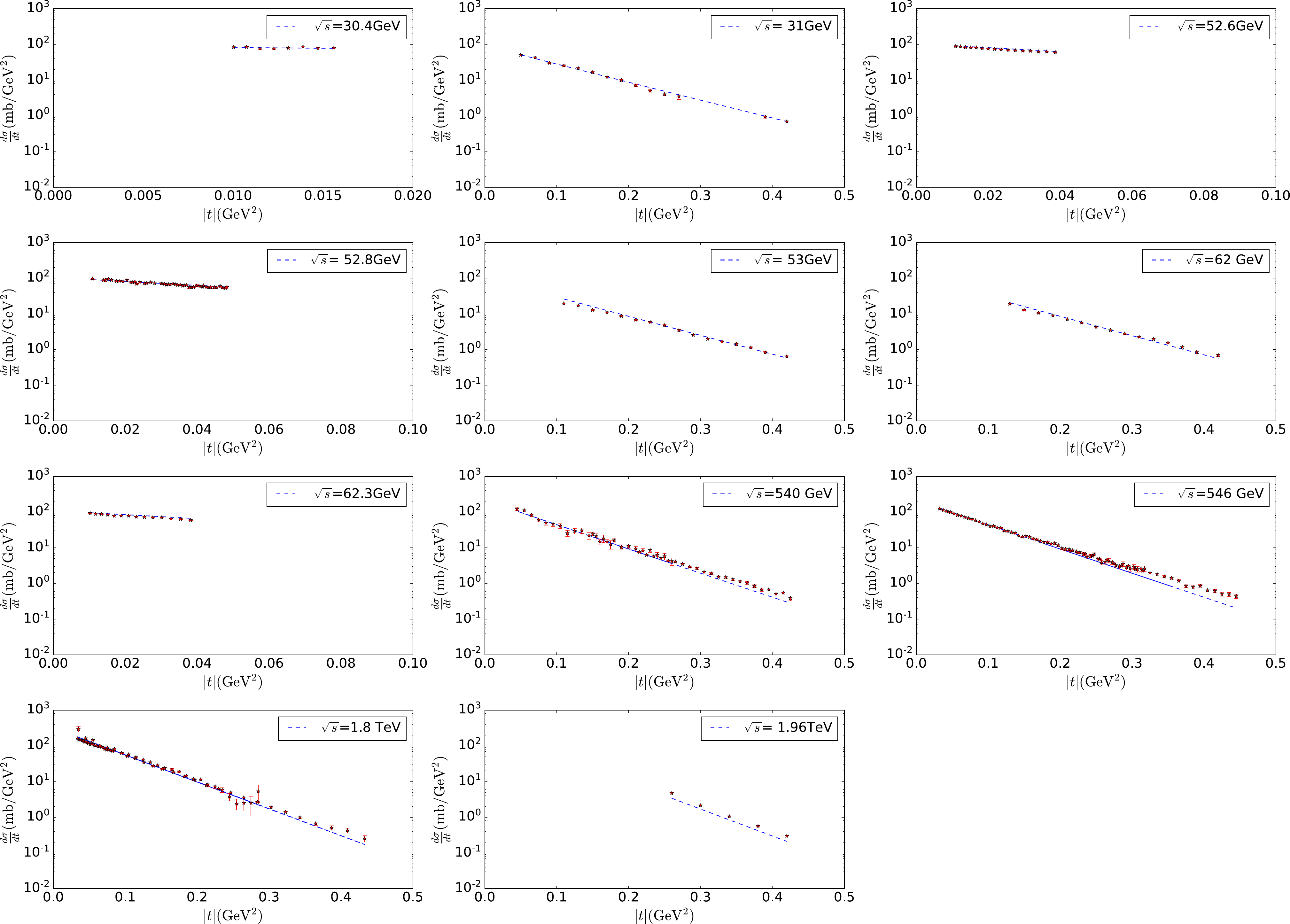}
\caption{
The differential cross section of the $p \bar{p}$ scattering %
as a function of $|t|$.
The dashed curves represent our calculations, %
and the experimental data are depicted by stars with error bars.
}
\label{DCS_ppbar}
\end{figure}

\begin{figure}[!tb]
\centering
\includegraphics[width=1.0\textwidth]{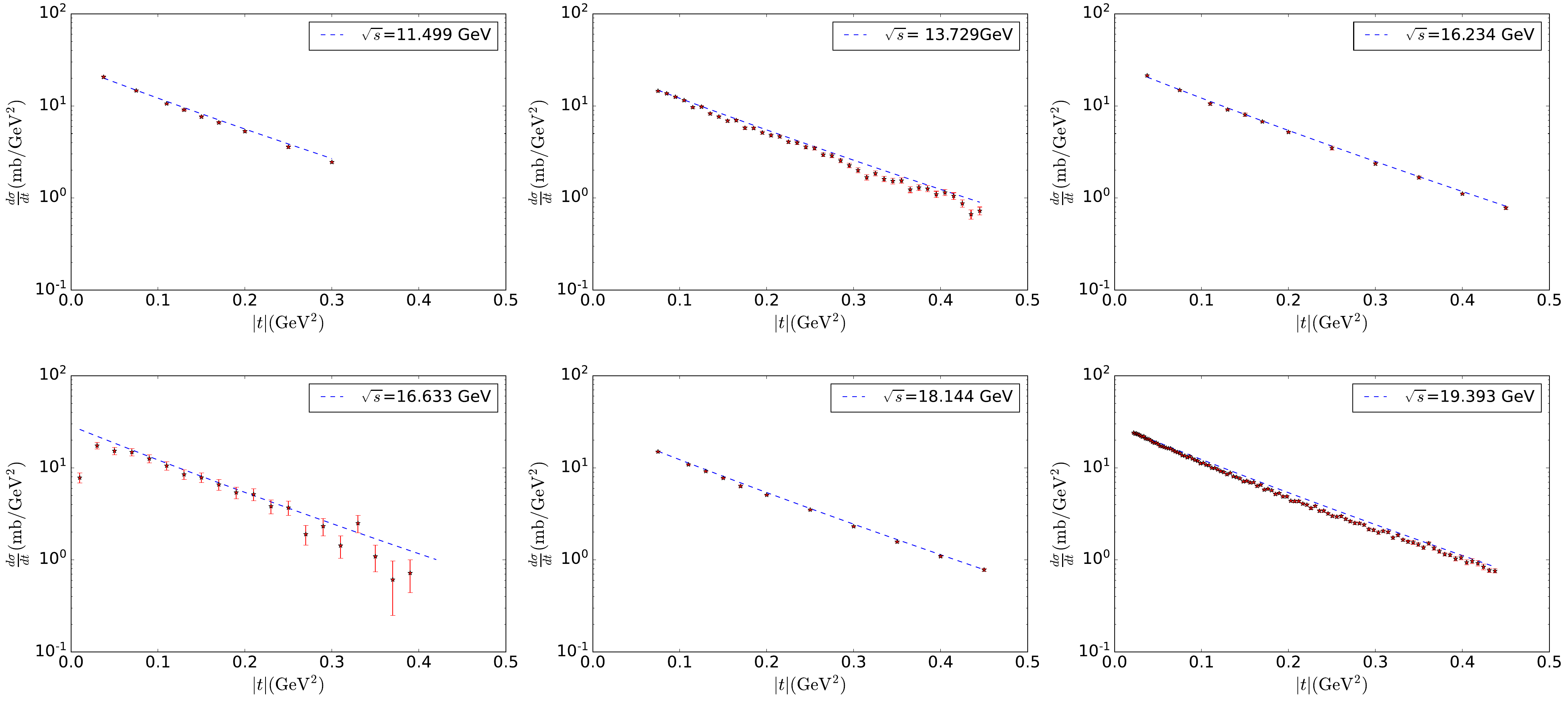}
\caption{
The differential cross section of the $\pi^+ p$ scattering %
as a function of $|t|$.
The dashed curves represent our calculations, %
and the experimental data are depicted by stars with error bars.
}
\label{DCS_piplus_p}
\end{figure}

\begin{figure}[!tb]
\centering
\includegraphics[width=1.0\textwidth]{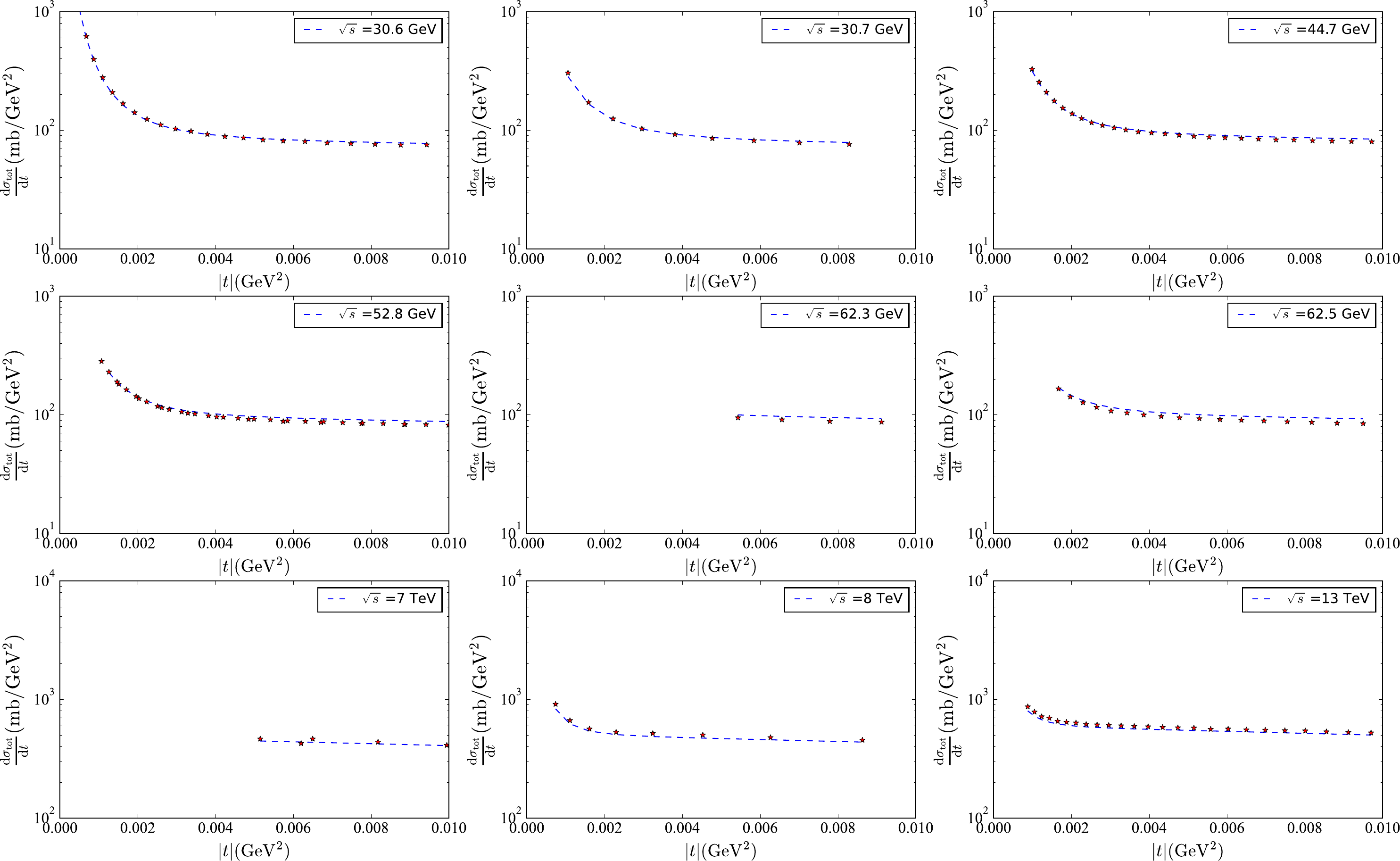}
\caption{
The total differential cross section of the $p p$ scattering %
as a function of $|t|$.
The dashed curves represent our calculations, %
and the experimental data are depicted by stars with error bars.
}
\label{DCS_C_pp}
\end{figure}

\begin{figure}[!tb]
\centering
\includegraphics[width=1.0\textwidth]{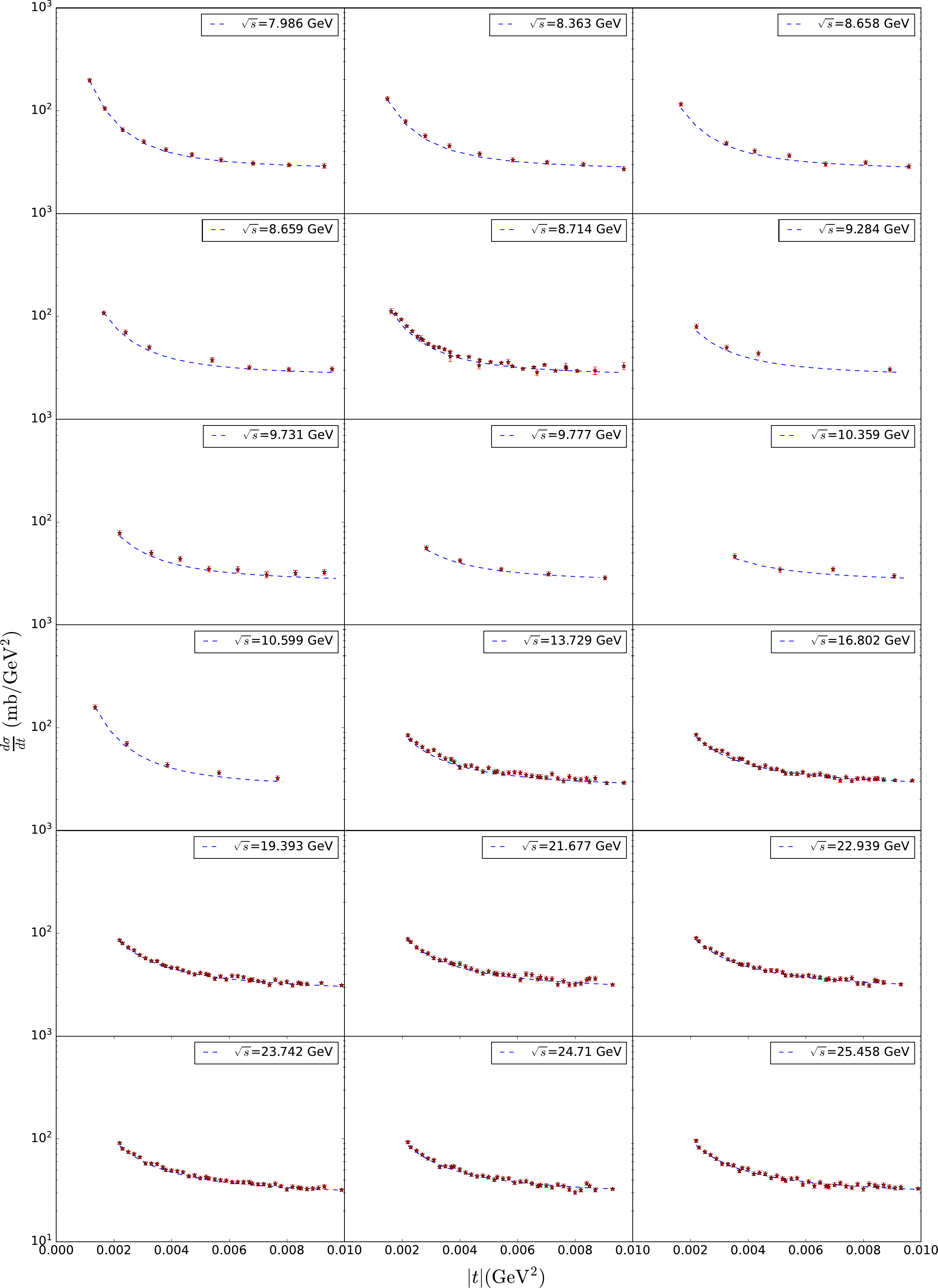}
\caption{
The total differential cross section of the $\pi^- p$ scattering %
as a function of $|t|$.
The dashed curves represent our calculations, %
and the experimental data are depicted by stars with error bars.
}
\label{DCS_C_piminus_p}
\end{figure}

Here we present our selected numerical results for the elastic $pp$, $p \bar{p}$, and $\pi p$ scattering.
Detailed explanations of the parameter determination and the obtained parameter values are presented in Refs.~\cite{Liu:2022zsa,Zhang:2023nsk,Liu:2023tjr,Zhang:2024psj}.
We display the results for the $pp$ and $p \bar{p}$ total cross section in Fig.~\ref{TCS_pp},
$\pi p$ total cross section in Fig.~\ref{TCS_pip},
$p \bar{p}$ differential cross section in Fig.~\ref{DCS_ppbar},
$\pi^+ p$ differential cross section in Fig.~\ref{DCS_piplus_p},
$p p$ total differential cross section (with the Coulomb interaction contribution) in Fig.~\ref{DCS_C_pp},
and $\pi^- p$ total differential cross section (with the Coulomb interaction contribution) in Fig.~\ref{DCS_C_piminus_p}.
From these results, it is found that the experimental data can be well described within the present model in the wide kinematic regions.

\label{sec:conclusion}
\section*{Conclusion}

We have investigated the total and differential cross sections of the elastic $pp$, $p \bar{p}$, and $\pi p$ scattering in the framework of holographic QCD, considering contributions of the Pomeron and Reggeon exchange in the Regge regime.
In our model setup, the Pomeron and Reggeon exchange are described by the Reggeized spin-2 glueball and vector meson propagator, respectively.
For the Pomeron-hadron (glueball-hadron) couplings, the gravitational form factors of the involved hadrons, which can be obtained with the bottom-up AdS/QCD models, are utilized.
Furthermore, for the differential cross sections in the very forward regions, contributions of the Coulomb interaction are also taken into account.

We have presented the obtained analytical expressions of the cross sections, and then explicitly shown the numerical results which are compared with the experimental data.
 For both the $pp (p \bar{p})$ and $\pi p$ cases, it has been seen that the data can be well described within the present framework.
Since this approach can be applied to other high energy scattering processes, further studies are certainly needed.
It is expected that our model will be tested at future experimental facilities and our understanding about the strong interaction and hadron structure will be further deepened.

\bibliographystyle{pepan}
\bibliography{references}

\end{document}